\documentstyle[12pt,epsfig]{article}

\newcommand{\be}{\begin{equation}}
\newcommand{\ee}{\end{equation}}

 1
 1
 1

\def\ref{\hang\noindent}

\def\la{\mathrel{\mathchoice {\vcenter{\offinterlineskip\halign{\hfil
$\displaystyle##$\hfil\cr<\cr\sim\cr}}}
{\vcenter{\offinterlineskip\halign{\hfil$\textstyle##$\hfil\cr<\cr\sim\cr}}}
{\vcenter{\offinterlineskip\halign{\hfil$\scriptstyle##$\hfil\cr<\cr\sim\cr}}}
{\vcenter{\offinterlineskip\halign{\hfil$\scriptscriptstyle##$\hfil\cr<\cr\sim\cr}}}}}
\def\ga{\mathrel{\mathchoice {\vcenter{\offinterlineskip\halign{\hfil
$\displaystyle##$\hfil\cr>\cr\sim\cr}}}
{\vcenter{\offinterlineskip\halign{\hfil$\textstyle##$\hfil\cr>\cr\sim\cr}}}
{\vcenter{\offinterlineskip\halign{\hfil$\scriptstyle##$\hfil\cr>\cr\sim\cr}}}
{\vcenter{\offinterlineskip\halign{\hfil$\scriptscriptstyle##$\hfil\cr>\cr\sim\cr}}}}}

\def\tem#1{\par\noindent
\hangindent6.5 mm\hangafter=0
\llap{#1\enspace}\ignorespaces}

\def \SAIT #1 #2 {{\em Mem.\ Soc.\ Astron.\ It.\/} {\bf #1}, #2}
\def \MESS #1 #2 {{\em The Messenger\/} {\bf #1}, #2}
\def \ASTRNACH #1 #2 {{\em Astron. Nach.\/} {\bf #1}, #2}
\def \AA #1 #2 {{\em Acta Astron.\/} {\bf #1}, #2}
\def \AAP #1 #2 {{\em Astron. Astrophys.\/} {\bf #1}, #2}
\def \AAL #1 #2 {{\em Astron. Astrophys. Lett.\/} {\bf #1}, L#2}
\def \AAR #1 #2 {{\em Astron. Astrophys. Rev.\/} {\bf #1}, #2}
\def \AAS #1 #2 {{\em Astron. Astrophys. Suppl. Ser.\/} {\bf #1}, #2}
\def \AJ #1 #2 {{\em Astron. J.\/} {\bf #1}, #2}
\def \ANNREV #1 #2 {{\em Ann. Rev. Astron. Astrophys.\/} {\bf #1}, #2}
\def \APJ #1 #2 {{\em Astrophys. J.\/} {\bf #1}, #2}
\def \APJL #1 #2 {{\em Astrophys. J. Lett.\/} {\bf #1}, L#2}
\def \APJS #1 #2 {{\em Astrophys. J. Suppl.\/} {\bf #1}, #2}
\def \APSS #1 #2 {{\em Astrophys. Space Sci.\/} {\bf #1}, #2}
\def \ASR #1 #2 {{\em Adv. Space Res.\/} {\bf #1}, #2}
\def \BAIC #1 #2 {{\em Bull. Astron. Inst. Czechosl.\/} {\bf #1}, #2}
\def \JSQRT #1 #2 {{\em J. Quant. Spectrosc. Radiat. Transfer\/} {\bf #1}, #2}
\def \MN #1 #2 {{\em Mon. Not. R. Astr. Soc.\/} {\bf #1}, #2}
\def \MEM #1 #2 {{\em Mem. R. Astr. Soc.\/} {\bf #1}, #2}
\def \PLR #1 #2 {{\em Phys. Lett. Rev.\/} {\bf #1}, #2}
\def \PASJ #1 #2 {{\em Publ. Astron. Soc. Japan\/} {\bf #1}, #2}
\def \PASP #1 #2 {{\em Publ. Astr. Soc. Pacific\/} {\bf #1}, #2}
\def \NAT #1 #2 {{\em Nature\/} {\bf #1}, #2}
\def \AA #1 #2 {{\em Acta Astron.\/} {\bf #1}, #2}
\def \SSREV #1 #2 {{\em Space Sci. Rev.\/} {\bf #1}, #2}
\def \aph #1 {{\em astro-ph\/} #1}
\def \IAUC #1 {{\em IAUC} #1} 

\begin{document}
\vspace*{1.8cm}
  \centerline{\bf BLACK HOLE CANDIDATES}
\vspace{1cm}
  \centerline{JANUSZ ZI\'O{\L}KOWSKI}
\vspace{1.4cm}
  \centerline{Copernicus Astronomical Center}
  \centerline{ul. Bartycka 18, 00-716 Warsaw, Poland}
\vspace{3cm}
\begin{abstract}
The state of the searches for isolated black holes, non-accreting black holes in binary systems and, finally, the accreting black holes in the X-ray binaries is presented. The third category is, by far, the most important source of information about the Galactic black holes and it is the main topic of this review. Different proposed signatures of an accreting black hole are reviewed (including the growing evidence for the presence of the event horizon in some systems). The list of 51 black hole candidates compiled with the help of these signatures is presented. To obtain a conclusive evidence that the candidate is really a black hole, one needs a dynamical mass estimate indicating that the mass of the compact object is greater than $\sim  3 M_\odot$. The observational methods of obtaining such estimates are briefly discussed. The application of these methods leads to the list of 18 objects that could be considered confirmed black holes. Their estimated masses are found to fall in the range $3.5 \div 16 M_\odot$. Finally, some general remarks concerning relation of black hole candidates to different classes of X-ray binaries such as X-ray Novae, microquasars and high/low mass X-ray binaries are presented. The implications of the properties of black hole binaries for the theory of stellar evolution are also discussed.
\end{abstract}

\vspace{2.0cm}

\section{Introduction}

Half a century ago, a black hole was just an exotic theoretical concept. Thirty years ago, the first Galactic black hole candidate (Cyg X$-$1) was proposed (Bolton, 1972). For the next ten years, Cyg X$-$1 remained the only candidate and some researchers put a lot of effort to devise a model of this system without invoking  a black hole. With the advent of the subsequent black holes in binary systems (LMC X$-$3, Mon X$-$1), the motivation for such efforts substantially diminished.  By today, black holes became a well established and extensively investigated class of observed astrophysical objects and their existence is no longer seriously disputed. Recently, searches for isolated Galactic black holes started and at least one object (microlens MACHO-99-BLG-22 = OGLE-1999-BUL-32) became a strong candidate. 

In the initial part of this article, I shall briefly describe the state of the searches for isolated black holes and for non-accreting black holes in binary systems. Most of my review will be devoted to black hole candidates in the X-ray binaries. I shall describe the methods of detecting black holes in such systems, review the properties of the candidates (51 objects) and discuss briefly some implications of these properties for the stellar evolution theory.

\section{Isolated Black Holes}

The evidence for detection of isolated black holes is much weaker than the one for binary black holes. However, some candidates were proposed among the objects responsible for long microlensing events and also among the objects listed in the Sloan Digital Sky Survey catalogues. It seems, that only one reasonably convincing case for an isolated black hole was demonstrated so far. 

\subsection{Extended Microlensing Events}

The long time scale of a microlensing event may be attributed either to a large mass of the lens or to the small relative transverse velocity of the lens with respect to the lensed object. The long events show, usually, the magnification fluctuations, reflecting the motion of the Earth. This effect permits to calculate the so called "microlensing parallax" which is a measure of the relative transverse motion of the lens with respect to the observer. Assuming standard model of the Galactic velocity distribution, one is then able to perform a likelihood analysis, which permits to estimate the distance and the mass of the lens. Basing on such analysis, several long events detected by the MACHO and OGLE projects were suggested as, possibly, caused by black hole lenses (Bennett et al. 2001, Mao et al. 2001, Bennett et al. 2002, Smith et al. 2002a). Agol et al. (2002) indicated that such analysis might overestimate the number of black hole lenses. At the moment, only one of the proposed candidates (MACHO-99-BLG-22 = OGLE-1999-BUL-32), with the time scale $\sim$  600 days) remains as a robust candidate for a black hole lens.

\subsection{Isolated Accreting Black Holes}

It has been estimated (Agol and Kamionkowski, 2001) that substantial number of isolated black holes may accrete from the interstellar medium with the efficiency sufficient to be detectable in deep X-ray surveys. Models of the emission from such objects suggest that they should also produce a synchrotron radiation detectable in the visible wavelengths (Ipser and Price 1982, McDovell 1985). Chisholm et al. (2002) searched through the 3.7 million objects in the Sloan Digital Sky Survey data and selected 150 000 objects with colors and properties consistent with synchrotron visible spectrum. 47 of these objects are ROSAT X-ray sources. For seven of these sources optical spectra excluding the candidacy (2 quasars and 5 F type stars) were observed. The remaining 40 sources are listed by the authors as possible black hole candidates.

\section{Non-Accreting Black Holes in Binary Systems}

Discovery of a black hole in a binary system with no mass transfer (and so no substantial X-ray emission) is rather difficult. Some W-R stars have unseen companions of several solar masses. However, these companions do not have to be compact objects. They   might be just normal main sequence stars, unseen because their luminosities are much lower than the luminosities of the bright W-R primaries. Cherepashchuk (1998) presents the list of 17 single-line W-R binaries, suspected of containing unseen compact components. Some of these components may be black holes, but no strong conclusions can be made.

\section{Signatures of an Accreting Black Hole in a Binary System}

Over the years, the researchers have accumulated a number of signatures, indicating that an accreting compact object in an X-ray binary is a likely black hole. These signatures could be ordered into several major classes:

$\bullet$ Signatures in the X-ray spectrum 

$\bullet$ Signatures in the X-ray variability 

$\bullet$ Level of the X-ray luminosity

$\bullet$ Evidence for the presence of the event horizon

\subsection{Signatures in the X-Ray Spectrum}

Over the last quarter of a century, a number of characteristics were proposed as X-ray signatures of an accreting black hole. Some of them, like {\bf flickering} or {\bf bimodal behaviour} ({\bf state transitions}) were found also in neutron star systems, so they are no longer believed to be unique properties of black holes. Those, listed below, should be considered as useful indicators (but not a definite proof) of the presence of a black hole.
\medskip
\tem{(1)} {\bf Presence of an ultrasoft component (kT$\sim$ 0.1$-$0.3 keV) in the X-ray spectrum}. This component, dominating in the high\footnote{"high" and "low" refer to X-ray flux below 10 keV, and not necessarily to the luminosity integrated over the entire X-ray range}  (soft) state, is believed to originate from the inner region of an accretion disc. This region has larger radius and lower temperature in the case of a black hole than in the case of a neutron star (see the point (4) below) and therefore produces softer X-ray spectrum.
\tem{(2)} {\bf Presence of a power law component (extending to several tens or several hundreds keV) in the X-ray spectrum}. This component, presumably produced through the up-comptonization of soft photons and sometimes described as an {\bf ultrahard component}, dominates in the low (hard) state. It is present also in some neutron star systems, but then it is not as hard as in black hole systems. In hard states, black holes are harder than neutron stars (Sunyaev, 2001).
\tem{(3)} {\bf Absence of the characteristic $\sim 2$ keV blackbody component at all levels of X-ray emission} (and, in particular, at high X-ray luminosity: $L_{\rm x} \ga 10^{37}$ erg s$^{-1}$). This component, prominent in the spectra of non-pulsating neutron star sources at high luminosity levels, is believed to originate on the surface of a weakly magnetized neutron star.

\noindent
Both the absence of the $\sim 2$ keV blackbody component and the presence of the supersoft component (see the point (1) above) contribute to the fact that black holes have softer X-ray colors than the neutron stars and, in the soft states, both classes of objects occupy different regions in the X-ray color-color diagram (Done and Gierli\'nski, 2002). They occupy different regions in the color-color diagram also in hard states, since then the power law component is harder in black holes than in neutron stars (point (2) above). 
\tem{(4)} {\bf Large inner radius of the disc and low color temperature of the inner edge of the disc}. These parameters are estimated from multi-color blackbody fits to the ultrasoft component of the X-ray spectrum. The results of such fits indicate, that the inner radii of discs around black hole candidates (BHCs) are, consistently, by a factor $\sim 4$ larger than for the discs around neutron stars (NSs) (Tanaka and Lewin, 1995). They indicate also, that the inner parts of discs around BHCs are cooler than the discs around NSs.

\subsection{Signatures in the X-Ray Variability}

Relatively recently, some characteristics of X-ray variability were proposed as signatures of an accreting black hole. They include:

\tem{(1)} {\bf Relatively narrow power density spectrum with high frequency cut-off at $\sim$ 10 Hz, during hard/low states}. Sunyaev and Revnitsev (2000) noticed that there are substantial differences between hard/low states power spectra of BHC and NS systems: (1) NSs have generally significantly broader  spectra  and (2) they demonstrate substantial power at high frequencies up to $\nu \ga$ 500 Hz, while BHCs have little power above $\nu \sim 10$-$50$ Hz.

\tem{(2)} {\bf Low frequency (0.1$-$3 Hz) quasi-periodic oscillations (QPOs)}. Both BHC and NS systems show low frequency QPOs. Generally, the low frequency QPOs of BHCs have lower frequencies (0.1$-$3 Hz), than those of NSs (5$-$60 Hz) (see e.g. Psaltis and Norman, 2000). However, it should be noted, that some BHCs show QPOs with frequencies $\sim$ 5$-$10 Hz (and so within NSs range).

\tem{(3)} {\bf High frequency quasi-periodic oscillations}. High frequency QPOs (or kHz QPOs) are seen both in NS systems (200 - 1330 Hz) and in BHC systems (41 - 450 Hz). There are different proposed mechanisms to explain their origin. At present, the leading contenders are: the parametric epicyclical resonance in the inner accretion disc (Abramowicz and Klu\'zniak, 2001) or the different modes of oscillations of the inner disc (e.g Wagoner et al., 2001). The fact that in some systems (and especially in some BHCs) the observed QPO frequencies form integral harmonics and that the frequencies seem to scale with the mass like $M^{-1}$ (see Remillard et al., 2002 for the most recent review) gives strong support to the first solution. In both theories, the values of the frequencies of the pair QPOs provide us with some constraints on the mass and the angular momentum of the compact object. Let us note, that (similarly as for low frequency QPOs) the frequencies are generally lower for BHCs than for NSs (presumably because of scaling with the mass - see Abramowicz et al., 2000). High frequency QPOs are therefore a good diagnostic tool to distinguish between accreting BHs and accreting NSs.

\subsection{Level of the X-Ray Luminosity}

A new class of point X-ray sources, called "ultraluminous compact X-ray sources" (ULXs) was established in recent years (Makishima et al., 2000). Objects of this type reside in many nearby galaxies (although none was found in our Galaxy or in M31). ULXs have properties of typical X-ray binaries, except that their X-ray luminosities are much higher: $L_{\rm x} \sim 10^{39} \div 10^{40}$ erg s$^{-1}$). This is too much for an accreting neutron star (unless it is a super-Eddington source). The most straightforward interpretation is that ULXs are accreting black holes with the masses $\sim 10 \div 1000 M_\odot$. However, the alternative explanation, assuming strongly collimated emission, is also possible and, so far, there is no consensus on this point.

\subsection{Evidence for the presence of the event horizon}

\tem{(1)} {\bf The lack of Type I bursts from some X-ray Novae}. X-ray Novae (XNe, called also Soft X-Ray Transients or SXTs) change their accretion rates (as inferred from the X-ray luminosities) by many orders of magnitude. Therefore, there must be time intervals, when the accretion rates fall into certain critical range of values, corresponding to the presence of the Type I bursts (thermonuclear flashes due to unstable nuclear burning of the accreted matter on the surface of the NS). About 50 XNe are known at present. In roughly 85 \% of them, the compact object is a BHC, in the remaining $\sim$ 15 \% it is a NS. In most of the NS XNe, Type I bursts are observed from time to time. However, not a single Type I burst was ever seen in any of the BHC XNe. This is a strong argument in favour of the presence of the event horizon, since Type I bursts can occur only if the solid surface is present. Narayan and Heyl (2002) investigated the stability of the nuclear burning on the surfaces of accreting compact objects of 1.4 and 10 $M_\odot$ for different accretion rates (assuming the presence of the solid surface in each case). In agreement with many earlier investigations, they found instability for some accretion rates in the case of 1.4 $M_\odot$ object (corresponding to the NS). But they found instability also for the 10 $M_\odot$ object. It seems, therefore, that independently on the mass, Type I bursts must appear, if only solid surface is present. The lack of these bursts implies, therefore, the presence of the event horizon around the compact object.

\tem{(2)} {\bf Very low quiescent luminosities of some X-ray Novae}. Between the outbursts, XNe retain some (although very low) level of X-ray luminosity, which indicates that some marginal accretion is still present. This, so called "quiescent luminosity" is so low, that it was not detectable, until the recent generation of the X-ray satellites. The comparison of the quiescent luminosities of BHC XNe and NS XNe shows a very substantial difference between the two classes of objects. For comparable orbital periods (which implies comparable sizes of the orbits and, so, the comparable accretion rates), the BHC XNe are, systematically, by about two - two and a half orders of magnitude dimmer than NS XNe (Narayan  et al. 2001, Garcia et al. 2001). The most obvious explanation of this difference is the presence of the solid surface (where most of the accretion energy is released) in the NS XNe. BHC XNe are "blacker" because they have event horizons.

\section{Black Hole Candidates in X-Ray Binaries}

Table 1 contains the list of BHCs in the X-ray binaries, compiled with the help of criteria discussed in the previous paragraph. I removed Cyg X-3 from the list, as there are huge (two orders of magnitude!) discrepancies in the estimates of its mass function (Schmutz et al., 1996, Hanson et al., 2000). However, I believe that Cyg X-3 will return to the list in not too distant future. The list contains, at present, 51 objects. Thanks to RXTE and Beppo-SAX it grew recently at a rate of 2$-$3 new BHCs per year.

I should stress that the signatures used to compile the list serve only as useful diagnostic tools: they help us to select systems suspected of harboring a black hole. They constitute, therefore, very useful criteria for compiling a list of {\it black hole candidates}. To obtain a conclusive evidence (and so, to include an object into the list of {\it confirmed black holes}), we have to obtain a dynamical mass estimate of the compact object and to demonstrate that this mass is too high for a neutron star.

\section{Confirmed Black Holes (Candidates with Dynamical Mass Estimates)}

\subsection{Upper Mass Limit for Neutron Stars}

\hspace{6mm}$\bullet$ {\bf Theory}

The maximum permitted mass of a neutron star (so called {\it Oppenheimer$-$Volkoff mass}) depends on the equation of state of matter at the densities higher than the nuclear density. Unfortunately, so far, this equation is not well known. The theory is not able to discriminate between different proposed models. For {\it seemingly reasonable} equations of state, the value of Oppenheimer$-$Volkoff mass $M_{\rm OV}$ ranges from $\sim 1.5 M_\odot$ for {\it soft} equations of state (density grows relatively fast with pressure) to $\sim 2.7 M_\odot$ for {\it hard} equations of state (density grows relatively slowly with pressure). If one assumes only general relativity as the correct theory of gravitation and the principle of causality (the speed of sound not greater than the speed of light) then, for arbitrarily exotic equation of state (constrained only by the requirement of the causality), one obtains $M_{\rm OV} \sim 3.2 M_\odot$ for non-rotating configuration and $M_{\rm OV} \sim 3.9 M_\odot$ for maximally rotating configuration. 

$\bullet$ {\bf Observations}

The most precise mass determinations are available for neutron stars that are members of binary radio pulsar systems composed of two neutron stars. The radio pulsar B1913+16 and its NS companion have measured masses 1.4411 and 1.3874 $M_\odot$, respectively, with the error $\pm$ 0.00035 $M_\odot$. For the system B1534+12 the masses of both components are 1.339 $M_\odot$ with the error $\pm$ 0.003 $M_\odot$. These two systems set the range of the observed values of the neutron star masses. There are numerous other determinations: three other binary NS systems, nine systems composed of radio pulsar and  white dwarf or main sequence star and eleven accreting X-ray pulsars (van Paradijs 1998, Thorsett and Chakrabarty 1999). The results range from 1.06 to 1.88 $M_\odot$, but the errors are much larger, so that there is no determination that would be inconsistent with the range established from the two systems mentioned above. The general rule is such that the more precise is the determination, the closer it approaches the value $\sim$ 1.4 $M_\odot$.

To summarize: all observational determinations of neutron star masses are consistent with the values in the range $1.34 \div 1.44 M_\odot$. 

$\bullet$ {\bf Conclusion}

If the accreting component of an X-ray binary has mass greater than $\sim  3 M_\odot$, it must be a black hole.

\subsection{Dynamical Mass Estimates}

There are several observational parameters used to estimate the mass of the compact component. They are briefly discussed below.

$\bullet$ {\bf The mass function}

This is the most important observational parameter used to constrain the mass of the compact component. The mass function $f(M_x)$ is calculated from the radial velocity variations of the optical companion:

$$f(M_x) = 1.0385 \times 10^{-7} K_{opt}^3P  \,M_\odot \eqno(1)$$

\noindent
where $K _{opt}$ is the semiamplitude of the radial velocities of the absorption lines of the optical component (in km/sec) and P is the orbital period (in days). The mass function is related to the masses of both components by:

$$f(M_x) = M_x^3 sin^3i/(M_{opt} + M_x)^2 \eqno(2)$$

\noindent
where $M_x$ and $M_{opt}$ are the masses of the compact and the optical components and $i$ is the inclination of the orbit. The value of $f(M_x)$ gives an absolute lower limit to the compact component mass. In many cases, this is enough to confirm the presence of a BH in the system (if $f(M_x)$ is $\ga  3 M_\odot$). Out of 18 BHs with the mass estimates, 7 systems have mass functions $\ga  5 M_\odot$ and 3 other systems have $f(M_x)$ in the range $3 \div 4 M_\odot$. However, in some cases, we need more precise constraints. To obtain them, we have to estimate the mass ratio $q = M_{opt}/M_x$ and the inclination of the orbit $i$.

$\bullet$ {\bf The rotational broadening of the absorption lines of the optical component}

From the measurements of the lines, we determine the projected rotational velocity at  the equator of the optical component $v_{rot} sin\, i$. Assuming the corotation of the optical component with the orbital motion (for a Roche lobe filling component, it is a very good assumption), we have:

$$ v_{rot} sin\, i = 0.46 K_{opt} (q/(1+q))^{1/3}  \eqno(3)$$

\noindent
With the help of this equation, one can determine the mass ratio $q$. The intrinsic width of the absorption lines is very small ( $\sim$ few km/sec), while the typical rotational broadening is of the order of few tens km/sec and can be, relatively easily, measured. This method has been applied, with a substantial success, to many BHC systems (see Orosz, 2002).

$\bullet$ {\bf  The radial velocities of the emission lines of the accretion disc}

The shifts of these lines reflect the orbital motion of the compact component and so permit us to determine the mass ratio directly:

$$ q = K_{em}/ K_{opt}   \eqno(4)$$

\noindent
where $K _{em}$ is the semiamplitude of the radial velocities of the emission lines. This method of determining the mass ratio is completely independent from the previous one. Unfortunately, it is rather uneasy to implement, since the emission lines are very broad ($\sim$ 2000 km/sec), while the orbital shifts are of the order of few tens km/sec.

$\bullet$ {\bf  The amplitude of the ellipsoidal light variations}

Due to filling of the Roche lobe, the optical component is tidally distorted and due to rotation it exhibits the ellipsoidal light variations (double sinusoid per orbital period if the rotation is synchronous). The amplitudes of these variations in V and I are given by:

$$ \Delta V = 0.26 sin^2 i /(1+q)  \eqno(5)$$

$$ \Delta I = 0.24 sin^2 i /(1+q)  \eqno(6)$$

\noindent
As may be seen, the dependence on the mass ratio $q$ is rather weak (for most of the BH XNe $q$ falls in the range 0.05 $\div$ 0.2) and, therefore, the ellipsoidal light variations provide us with a valuable information about the inclination $i$ (even, if we do not know $q$). In practice, the procedure is not that simple, since the optical light is usually (even in the quiescence) contaminated by the residual contribution from the accretion disc. Using light curves in different colors, one obtains different values for the inclination (see the discussion of A 0620$-$00 in section 6.3). To make the situation worse, this residual contribution is frequently variable. Using the observational data from different time intervals, one again obtains different values for the inclination. Extraction of the true value requires, often, a very careful modeling (Froning and Robinson 2001, Gelino et al. 2001b).

$\bullet$ {\bf  The mass-spectral type relation for the optical component}

The typical optical components of BHXNe are the lower main sequence stars, which satisfy reasonably well the mass-spectral type relation. This relation may be used to estimate the mass of the compact component, if the mass ratio is unknown. Since the mass of the optical component is usually quite small (below 1 $M_\odot$), even substantial uncertainty (say factor of two) does not influence significantly the mass estimate for the compact component.

$\bullet$ {\bf  High frequency QPOs}

As discussed in section 4.2 p. (3), the sets of frequencies of the kHz QPOs provide us with some information on the mass of the compact object.

\subsection{Black Holes Candidates with Dynamical Mass Estimates}

The dynamical mass estimates, indicating the mass of the compact component greater than $\sim 3 M_\odot$, are now available for 18 X-ray binaries. The estimated masses of black holes in these 18 systems are given in Table 1.
 These estimates are based on the literature and discussed briefly below. Recently, Orosz (2002) performed new calculations of all masses using the published values of mass functions, inclinations and rotational velocities and employing a simple Monte Carlo code. Such approach has an advantage of a relatively uniform treatment of all systems and a disadvantage of omitting important pieces of information considered by different authors while discussing the individual systems.

Below, I shall briefly discuss the parameters of these 18 systems.

\tem{(1)} {\bf Cyg X$-$1}

At present, there are no doubts about the presence of black hole in the system. However, in spite of three decades of investigations, there is still substantial uncertainty concerning the masses of both components. The value of the mass function given by Gies and Bolton (1982): $f(M_x) = 0.25 \pm 0.01 M_\odot$ was only slightly modified recently by Brocksopp et al. (1999): $f(M_x) = 0.244 \pm 0.005 M_\odot$. Gies and Bolton (1986a, 1986b), analyzing the emission lines of the stellar wind,  the rotational broadening of the absorption lines of the optical component and the photometric V-band light curve, found that this component must be close to filling its Roche lobe (the filling factor must be greater than 0.9). From the same analysis, they estimated the inclination of the orbit to be $i = 33^{\rm o} \pm 5^{\rm o}$. From polarimetric measurements (in three colors) Dolan and Tapia (1989) found a rather wide range of values for inclinations: $i = 62^{\rm o}  (+ 5^{\rm o}, - 37^{\rm o})$. Paczy\'nski (1974) has shown that basing only on
$\bullet$ {\it mass function} $\bullet$ {\it lack of X-ray eclipses} $\bullet$ {\it obvious fact that the optical component cannot be larger than its Roche lobe} and $\bullet$ {\it the spectral type and the photometry of the optical component}, one can obtain lower limits to the masses of both components as functions of the distance to the system. Unfortunately, this distance is not well established. However, most likely, Cyg X$-$1 belongs to Cyg OB$-$3 association and so its distance is d = 2.15 $\pm$ 0.07 kpc (Massey et al., 1995). With this distance, the present day calibrations of effective temperatures and bolometric corrections: $T_e$ = 32000 K (Herrero et al., 1995), B.C. = -3.18 (Vacca et al., 1996) and Massey's photometry, following Paczy\'nski's reasoning, one obtains $\sim 8 M_\odot$ and $\sim 28 M_\odot$ as lower limits to the masses of the compact and the optical components, respectively (Zi\'o{\l}kowski, 2002).

 Quite independent constraints on the masses may be obtained from analysis of the evolutionary status of the optical component (Zi\'o{\l}kowski, 2002). At the distance of 2.15 $\pm$ 0.07 kpc, its luminosity must be $M_{\rm bol} \approx - 9.4 \pm 0.2$. It can be argued that it must be a core hydrogen burning configuration and then its mass cannot be smaller than $\sim 31 \pm 3 M_\odot$. This lower limit is consistent with the values discussed above. Even if the distance to the system was as small as  1.8 kpc (the smallest value quoted in the literature), then still the optical component should be more massive than $\sim 24$ $M_\odot$. Altogether, the evolutionary analysis indicates that the mass of HDE 226868 is probably in the range $30 \div 40$ $M_\odot$ and the initial mass of its progenitor was probably $45 \div 55$ $M_\odot$. The mass of the compact component is probably in the range $15 \div 17$ $M_\odot$.  These results values are fully consistent with the values $M_{\rm opt} = 33 \pm 9 M_\odot$ and $M_{\rm x} = 16 \pm 5 M_\odot$, obtained by Gies and Bolton (1986a) from the essentially distance independent (but model dependent) analysis.

Orosz (2002) obtained $M_{\rm x} = 6.85 \div 13.25 M_\odot$. 

\tem{(2)} {\bf LMC X$-$3}

The mass function is given by Cowley et al. (1983): $f(M_x) = 2.3 \pm 0.3 M_\odot $. The same  authors estimated the rotational broadening of the absorption lines of the optical component to be 130 $\pm$ 20 km/sec and concluded that the inclination of the orbit must be limited to $i \sim 50^{\rm o} \div 70^{\rm o}$ (the upper limit comes from the lack of X-ray eclipses). Their estimate of $M_x$ was $7 \div 14  M_\odot $ with the most likely value $\sim 9 M_\odot $. Kuiper et al. (1988) included the photometric data (exhibiting the ellipsoidal light variations) and obtained $i = 64^{\rm o} \div 70^{\rm o}$ and $M_x = 4.5 \div 6.5 M_\odot $.

Orosz (2002), using rotational velocity of Cowley et al. and inclination of Kuiper et al., obtained $M_{\rm x} = 5.94 \div 9.17 M_\odot$. 

\tem{(3)} {\bf LMC X$-$1}

The mass function, given by Hutchings et al. (1987), is $f(M_x) = 0.14 \pm 0.05 M_\odot $. The lack of X-ray eclipses implies $i \la 60^{\rm o}$. Hutchings et al. obtain $M_x \sim 4 \div 10 M_\odot $ for the mass of the compact component. Shrader and Titarchuk (1999) argue for much higher value ($M_x \ga 16 M_\odot $), basing on their modeling of the accretion disc X-ray spectrum. It is not clear how to reconcile these two estimates. Orosz (2002) quotes the value of Hutchings et al.

\tem{(4)} {\bf SS 433}

This celebrated object has been extensively reviewed on many occasions (see e.g. Margon, 1984). Here, I shall consider only the data important for the mass estimates.

The system displays two basic periods: the binary orbital period ($P_{\rm orb} = 13.087 \pm 0.003$ d) and the period of the precession of the accretion disc and jets ($P_{\rm prec} \approx 162.5$ d). The inclination of the orbital plane (from the kinematic model of the jets) is $i = 78.82^{\rm o} \pm 0.11^{\rm o}$, the precession angle $\theta = 19.80^{\rm o} \pm 0.18^{\rm o}$, the distance to the object (from the radio observations of the jets) $d \approx 5$ kpc. The visual extinction is estimated to be $A_{\rm V} \sim 8$ magnitudes. The system is very luminous in the optical band: $L_{\rm opt} \geq 2 \times 10^{38}$ erg s$^{-1}$. About 80 \% of this luminosity comes from the photosphere of geometrically thick accretion disc, which can be treated as a flattened star. The observed He II emission lines also originate on the surface of the disc and reflect its orbital motion. Until recently, no spectral lines produced by the second member of the binary (the mass donor or the "optical component") were seen. Therefore, untypically, only the mass function $f(M_{\rm opt})$ could be estimated. Unfortunately, its value is not established very well. Looking through the literature, one can find $f(M_{\rm opt}) \approx 10.1 M_\odot $ (Crampton and Hutchings, 1981), $f(M_{\rm opt}) = 7.7 (+ 3.0, - 2.4) M_\odot$ (Fabrika and Bychkova, 1990) and $f(M_{\rm opt}) = 1.90 \pm 0.25M_\odot$ (D'Odorico et al., 1991). Fabrika and Bychkova gave the most convincing arguments supporting their value and, therefore, it will be used in further considerations.
Substantial efforts were undertaken to derive the mass ratio $Q = M_{\rm x}/ M_{\rm opt}$ from the modeling of the optical light curves. The light curves in two colors are two-dimensional (B and V magnitudes are modulated with two periods: $P_{\rm orb}$ and $P_{\rm prec}$) and rather complicated. The modeling assumes mutual partial eclipses of two photospheres (that of the accretion disc and of a "normal star", which must be a massive early type star) with possible addition of a hot spot on the surface of the disc. Two such analysis carried out by Leibowitz (1984) and Antokhina and Cherepashchuk (1985), under different assumptions about the geometrical shape of the disc gave very similar results. Both papers got thickness of the disc equal approximately to its radius (ratio of polar to equatorial radii of about 0.5). Leibowitz obtained $Q \geq 0.8$ and the preferable value of  $Q \sim 1.2$. Antokhina and Cherepashchuk got acceptable solutions for $Q \approx 0.4$ and $Q \ga 1.1$ (they preferred Q $\sim$ 1.2). Quantitatively, the results of these modeling were very unprecise (with the mass function of Fabrika and Bychkova, the value of $M_{\rm x}$ is in the range $4.2 \div 62 M_\odot$), but they confirmed the presence of a black hole in the system.

Quite recently, Gies et al. (2002) observed, finally, the absorption lines of the optical component (mass donor). They estimated the spectrum as A-type supergiant and measured the semiamplitude of the radial velocities: $K_{opt} = 100 \pm 15$ km/sec (which corresponds to $f(M_x) = 1.36 (+ 0.71, - 0.52) M_\odot $. Taking the semiamplitude of the radial velocities of the emission lines from Fabrika and Bychkova: $K_x = 175 \pm 20$ km/sec, they estimate the mass ratio $M_{\rm x}/M_{\rm opt} = 0.57 \pm 0.11$ and the masses $M_{\rm opt} = 19 \pm 7 M_\odot$ and $M_{\rm x} = 11 \pm 5 M_\odot$.  

\tem{(5)} {\bf GRO J0422+32}

The most recent values of the mass function, are given by Harlaftis et al. (1999): $f(M_x) = 1.13 \pm 0.09 M_\odot $ and Webb et al. (2000): $f(M_x) = 1.191 \pm 0.021 M_\odot $. Filippenko et al. (1995) estimated the mass ratio from the radial velocities of the accretion disc emission lines and got $q = 0.109 \pm 0.009$. This value is consistent with the ones obtained from the rotational broadening of the absorption lines of the optical component: $q = 0.116 (+0.079, -0.071)$ (Harlaftis and Filippenko, 2000) and $q = 0.111 (+ 0.048, -0.022)$ (Webb et al., 2000). Taking into account $f(M_x)$ and $q$ and assuming normal (for its spectral type) mass of the secondary ($\sim 0.4 M_\odot$) Filippenko et al. (1995) obtained $i = 48^{\rm o} \pm 3^{\rm o}$ and $M_{\rm x} = 3.57 \pm 0.34 M_\odot$.

Orosz (2002) obtained $M_{\rm x} = 3.66 \div 4.97 M_\odot$. 

\tem{(6)} {\bf A 0620$-$00}

The mass function is $f(M_x) = 2.72 \pm 0.06 M_\odot $ (Marsh et al., 1994). The same authors used the absorption lines of the optical component to estimate the mass ratio: $q = 0.067 \pm 0.010$. Mass ratio was estimated also from the radial velocities of the accretion disc emission lines: Haswell and Shafter (1990) obtained $q = 0.094 (+0.022, -0.015)$, while Orosz et al. (1994) got $q = 0.074 \pm 0.006$. The determinations of the inclination of the orbit, estimated from the ellipsoidal light variations have a long and complicated history. Haswell et al. (1993), using variations in the UBVR bands, found $i = 70^{\rm o} \pm 3.5^{\rm o}$, which lead to the mass of the compact component $M_x = 3.8 \div 4.4 M_\odot $. Shahbaz et al. (1994a) used the variations in the JK bands and got $i = 37^{\rm o} \pm 5^{\rm o}$, which lead to $M_x = 14 \pm 7 M_\odot $. The second determination seemed to be closer to the truth, since the flux in the infrared bands should be less contaminated by the residual contribution of the accretion disc. However, the next determination, using the H band photometry (Froning and Robinson, 2001), did not solve the problem. The authors found, first, that the light curves change from year to year and, second, that the determination of $i$ is strongly model dependent and admits a wide range of values for the inclination. They decided only to conclude that $i$ should be in the range $38^{\rm o} \div 75^{\rm o}$, which corresponds to $M_x = 3.3 \div 13.6 M_\odot $. Finally, Gelino et al. (2001b) succeeded in modeling their new JHK photometry (they had to introduce cold spots on the surface of the optical component) and found $i = 40.8^{\rm o} \pm 3^{\rm o}$ and $M_x = 11.0 \pm 1.9 M_\odot $.
 
Orosz (2002) obtained $M_{\rm x} = 8.70 \div 12.86 M_\odot$. 

%\vspace{2.0cm} 
%\pagebreak
%\vspace{1cm} %TO ALLOW SUFFICIENT SPACE BETWEEN THE TEXT AND THE FIGURES
%\begin{table}
\centerline{\bf Tab. 1 $-$ Black Hole Candidates in X-Ray Binaries}
\nobreak
\vspace{5mm} 
%\nopagebreak

\moveleft 12mm
\vbox{
\begin{tabular}{|rcl|l|l|l|r|r|c|r|}
\hline
&&&&&&&&&\\
\multicolumn{3}{|c|}{Name}&\multicolumn{1}{|c|}{P$_{orb}$}&\multicolumn{1}{|c|}{Opt. Sp} &\multicolumn{1}{|c|}{Other names}&\multicolumn{1}{ |c|}{X$-$R}&\multicolumn{1}{|c|}{C}&\multicolumn{1}{|c|}{M$_{\rm BH}$/ M$_\odot$}& \multicolumn{1}{|c|}{Ref}\\
&&&&&&&&&\\
\hline
&&&&&&&&&\\
Cyg X\hspace*{-2.4ex}&$-$&\hspace*{-2.4ex}1&5$^d$6&O9.7 Iab&HDE 226868(O)&pers&$\mu$Q&16 $\pm$ 5&\\
&&&&&V1357 Cyg (O)&&&&\\
LMC X\hspace*{-2.4ex}&$-$&\hspace*{-2.4ex}3&1$^d$70&B3 V&&pers&&6 $\div$ 9&\\
LMC X\hspace*{-2.4ex}&$-$&\hspace*{-2.4ex}1&4$^d$22&O7$-$9 III&&pers&&4 $\div$ 10&\\
SS\hspace*{-2.4ex}&&\hspace*{-2.4ex}433&13$^d$1&$\sim$ A I&V1343 Aql (O)&pers&$\mu$Q&11 $\pm$ 5&\\
GX 339\hspace*{-2.4ex}&$-$&\hspace*{-2.4ex}4&14$^h$8&F8$-$G2 III&V821 Ara (O)&RT& &&1,2,3\\
3U 0042\hspace*{-2.4ex}&+&\hspace*{-2.4ex}32&11$^d$6&G ?&&T&&&4\\
XTE J0421\hspace*{-2.4ex}&+&\hspace*{-2.4ex}560&&B[e] I&CI Cam (O)&T&$\mu$Q?&&5,6\\
GRO J0422\hspace*{-2.4ex}&+&\hspace*{-2.4ex}32&5$^h$09&M2 V&V518 Per (O)&T&&3.6 $\div$ 5.0&\\
&&&&&XRN Per 1992&&&&\\
A 0620\hspace*{-2.4ex}&$-$&\hspace*{-2.4ex}00&7$^h$75&K4 V&V616 Mon (O)&RT&&11 $\pm$ 2&\\
&&&&&Mon X$-$1&&&&\\
&&&&&XN Mon 1975&&&&\\
GRS 1009\hspace*{-2.4ex}&$-$&\hspace*{-2.4ex}45&6$^h$96&K8 V&MM Vel (O) &T&&4.4 $\div$ 4.7&\\
&&&&&XN Vel 1993&&&&\\
XTE J1118\hspace*{-2.4ex}&+&\hspace*{-2.4ex}480&4$^h$1&K7$-$M0 V&KV Uma (O)&T&&$6.0 \div$ 7.7&\\
GS 1124\hspace*{-2.4ex}&$-$&\hspace*{-2.4ex}684&10$^h$4&K0$-$5 V&GU Mus (O)&T&&7.0 $\pm$ 0.6&\\
&&&&&XN Mus 1991&&&&\\
GS 1354\hspace*{-2.4ex}&$-$&\hspace*{-2.4ex}645&&+&BW Cir (O)&T&&&\\
&&&&&Cen X$-$2 ?&&&&7\\
A 1524\hspace*{-2.4ex}&$-$&\hspace*{-2.4ex}617&&+&KZ TrA (O)&RT&&&\\
&&&&&TrA X$-$1&&&&7\\
4U 1543\hspace*{-2.4ex}&$-$&\hspace*{-2.4ex}475&1$^d$12&A2 V&IL Lup (O)&RT&&8.4 $\div$ 10.4&\\
XTE J1550\hspace*{-2.4ex}&$-$&\hspace*{-2.4ex}564&1$^d$55&G8 IV$-$K4 III&V381 Nor (O)&RT&$\mu$Q&9.7 $\div$ 11.6&\\
4U 1630\hspace*{-2.4ex}&$-$&\hspace*{-2.4ex}472&&+(IR)&Nor X$-$1&RT&&&4,7,8\\
XTE J1650\hspace*{-2.4ex}&$-$&\hspace*{-2.4ex}500&0$^d$212&G$-$K&&T&$\mu$Q &&9,10\\
GRO J1655\hspace*{-2.4ex}&$-$&\hspace*{-2.4ex}40&2$^d$62&F3$-$6 IV& V1033 Sco (O)&RT&$\mu$Q&6.3 $\pm$ 0.3&\\
&&&&&XN Sco 1994&&&&\\
H 1705\hspace*{-2.4ex}&$-$&\hspace*{-2.4ex}250&12$^h$5&K5 V&V2107 Oph (O)&T&&5.7 $\div$ 7.9&\\
&&&&&XN Oph 1977&&&&\\
XTE J1709\hspace*{-2.4ex}&$-$&\hspace*{-2.4ex}267&&&&T&&&11\\
GRO J1719\hspace*{-2.4ex}&$-$&\hspace*{-2.4ex}24&&M0$-$5 V&V2293 Oph (O)&T&&&3,12\\
&&&&&XN Oph 1993&&&&\\
XTE J1720\hspace*{-2.4ex}&$-$&\hspace*{-2.4ex}318&&&&T&&&13\\
KS 1730\hspace*{-2.4ex}&$-$&\hspace*{-2.4ex}31&&&&T&&&12\\
GRS 1737\hspace*{-2.4ex}&$-$&\hspace*{-2.4ex}31&&&&T&&&12\\
GRS 1739\hspace*{-2.4ex}&$-$&\hspace*{-2.4ex}278&&$\ga$ F5 V&&T&&&3,14\\
XTE J1739\hspace*{-2.4ex}&$-$&\hspace*{-2.4ex}302&&&&T&&&15\\
1E 1740.7\hspace*{-2.4ex}&$-$&\hspace*{-2.4ex}2942&12$^d$73&&&pers&$\mu$Q&&12,16\\
H 1741\hspace*{-2.4ex}&$-$&\hspace*{-2.4ex}322&&&&T&&&1\\
H 1743\hspace*{-2.4ex}&$-$&\hspace*{-2.4ex}32&&&&T&&&12\\
SLX 1746\hspace*{-2.4ex}&$-$&\hspace*{-2.4ex}331&&&&T&&&12\\
\end{tabular}}
%\end{table}
%\newpage

%\begin{table}
%\vspace{7mm}
\centerline{\bf Tab. 1 $-$ Black Hole Candidates in X-Ray Binaries (continued)}
\nobreak
\vspace{5mm}
  
\moveleft 6mm
\vbox{
\begin{tabular}{|rcl|l|l|l|r|r|c|r|}
\hline
&&&&&&&&&\\
\multicolumn{3}{|c|}{Name}&\multicolumn{1}{|c|}{P$_{orb}$}&\multicolumn{1}{|c|}{Opt. Sp} &\multicolumn{1}{|c|}{Other names}&\multicolumn{1}{ |c|}{X$-$R}&\multicolumn{1}{|c|}{C}&\multicolumn{1}{|c|}{M$_{\rm BH}$/ M$_\odot$}& \multicolumn{1}{|c|}{Ref}\\
&&&&&&&&&\\
\hline
&&&&&&&&&\\
XTE J1748\hspace*{-2.4ex}&$-$&\hspace*{-2.4ex}288&&&&T&$\mu$Q&&17\\
4U 1755\hspace*{-2.4ex}&$-$&\hspace*{-2.4ex}338&4$^h$4 ?&+&V4134 Sgr (O)&pers&&&12\\
XTE J1755\hspace*{-2.4ex}&$-$&\hspace*{-2.4ex}324&&&&T&&&18\\
GRS 1758\hspace*{-2.4ex}&$-$&\hspace*{-2.4ex}258&18$^d$45&&&pers&$\mu$Q&&12,16,17\\
SAX J1805.5\hspace*{-2.4ex}&$-$&\hspace*{-2.4ex}2031&&&&T&&&19,20\\
XTE J1806\hspace*{-2.4ex}&$-$&\hspace*{-2.4ex}246&&&&T&&&21\\
XTE J1819\hspace*{-2.4ex}&$-$&\hspace*{-2.4ex}254&2$^d$817&B9 III&V4641 Sgr (O)&T&$\mu$Q&$6.8 \div 7.4$&\\
RX J1826.2\hspace*{-2.4ex}&$-$&\hspace*{-2.4ex}1450&4$^d$117&O6.5 Vf&LS 5039 (O)&pers&$\mu$Q&&22,23\\
EXO 1846\hspace*{-2.4ex}&$-$&\hspace*{-2.4ex}031&&&&T&&&12\\
XTE J1856\hspace*{-2.4ex}&+&\hspace*{-2.4ex}053&&&&T&&&14\\
XTE J1859\hspace*{-2.4ex}&+&\hspace*{-2.4ex}226&9$^h$16&$\sim$ G 5&V404 Vul (O)&T&&$8 \div 10$&\\
XTE J1901\hspace*{-2.4ex}&+&\hspace*{-2.4ex}014&&&&RT&&&24\\
XTE J1908\hspace*{-2.4ex}&+&\hspace*{-2.4ex}094&&+(IR)&&RT&&&25\\
GRS 1915\hspace*{-2.4ex}&+&\hspace*{-2.4ex}105&33$^d$5&K$-$M III&V1487 Aql (O)&RT&$\mu$Q&14 $\pm$ 4&\\
&&&&&XN Aql 1992&&&&\\
4U 1918\hspace*{-2.4ex}&+&\hspace*{-2.4ex}146&&&&T&&&4\\
4U 1957\hspace*{-2.4ex}&+&\hspace*{-2.4ex}115&9$^h$3&+&V1408 Aql (O)&pers&&&12\\
GS 2000\hspace*{-2.4ex}&+&\hspace*{-2.4ex}251&8$^h$3&K5 V& QZ Vul (O)&T&&7.1 $\pm$ 7.8&\\
&&&&&XN Vul 1988&&&&\\
XTE J2012\hspace*{-2.4ex}&+&\hspace*{-2.4ex}381&&&&T&&&26,27\\
GS 2023\hspace*{-2.4ex}&+&\hspace*{-2.4ex}338&6$^d$46&K0 IV& V404 Cyg (O)&RT&&10.0 $\div$ 13.4&\\
&&&&&XN Cyg 1989&&&&\\
2S 2318\hspace*{-2.4ex}&+&\hspace*{-2.4ex}62&&&&T&$\mu$Q&&28\\
&&&&&&&&&\\
\hline
\end{tabular}}

%\end{table}
\vspace{10mm}  
\nopagebreak

{\small NOTES:\vspace{2mm}\\
%\vspace{4mm}  
P$_{orb}$ $-$ orbital period\\
Opt. Sp $-$ optical spectrum\\
X-R $-$ X-ray variability\\
C $-$ comments\\
M$_{\rm BH} -$  mass of black hole component\\
Ref $-$ references\\
$+$  $-$ optical counterpart was identified, but the spectrum was not obtained\\ 
IR $-$ optical counterpart was seen only in infrared\\
(O) $-$ name of the optical object\\
T $-$ transient\\
RT $-$ recurrent transient\\
pers $-$ persistent\\
$\mu$Q $-$ microquasar\\}
  
{\small REFERENCES:\vspace{2mm}\\
(1) Tanaka and Lewin, 1995; Zdziarski et al., 1998; (3) Chaty et al., 2001; (4) Chen et al., 1997; (5) Orlandini et al., 2000; (6)Robinson et al., 2002; (7) Liu et al., 2001; (8) Augusteijn et al., 2001; (9) Tomsick et al., 2002; (10) Sanchez-Fernandes et al., 2002; (11) Marshall et al., 1997; (12) van Paradijs,  1998; (13) Remillard et al., 2003; (14) Remillard, 1999; (15) Smith et al., 1998; (16) Smith et al., 2002b; (17) Mirabel, 2000; (18) Remillard et al., 1997; (19) Lowes et al., 2002; (20) Markwardt et al., 2002;  (21) Hynes et al., 1998; (22) Paredes et al., 2000; (23) McSwain and Gies, 2002; (24) Remillard and Smith, 2002; (25) Chaty and Megani, 2002; (26) Remillard et al., 1998; (27) Callanan et al., 1998; (28) Taylor et al., 1991.

References for confirmed candidates are given in section 6.3}

\vspace{8mm}

\tem{(7)} {\bf GRS 1009$-$45}

The mass function, based on observations with the help of Keck II telescope, is $f(M_x) = 3.17 \pm 0.12 M_\odot$ (Filippenko et al., 1999). The same observations yielded precise measurements of the radial velocities of the accretion disc emission lines, which permitted to calculate the mass ratio $q = 0.137 \pm 0.015$. The lack of X-ray eclipses implies $i \la 80^{\rm o}$. Assuming that the secondary has a normal mass for its spectral type ($\sim 0.6 M_\odot$), the authors obtain $M_{\rm x} \approx 4.4 M_\odot$ and $i \approx 78^{\rm o}$. The mass of the compact object cannot be smaller because of the limits set by the lack of X-ray eclipses. It can be somewhat larger (up to $\sim 4.7 M_\odot$ if the mass of the secondary is $\sim 0.75 M_\odot$). 

Orosz (2002) quotes, after Filippenko et al. (1999), $M_{\rm x} = 3.64 \div 4.74 M_\odot$. 

\tem{(8)} {\bf XTE J1118+480}

The most recent value of mass function, given by Wagner et al. (2001) is $f(M_x) = 6.1 \pm 0.3 M_\odot$. McClintock et al. (2001) found ellipsoidal light variations in the I band with the amplitude $\sim 0.15$  mag. Assuming that the secondary has a normal mass for its spectral type ($\sim 0.5 M_\odot$), and considering different possible values of the disc contribution to the I-band emission (40\% to 66\%), McClintock et al. found $i = 52^{\rm o} \div 80 ^{\rm o}$ and $M_{\rm x} = 7.2 \div 13.2 M_\odot$. Wagner et al. (2001) analyzed their R band photometry and found ellipsoidal light variations with the amplitude $\sim 0.2$  mag. Modeling their light curves and permitting variable contribution from the disc (decreasing with time from 72\% to 64\%), they found $i = 81^{\rm o} \pm 2 ^{\rm o}$ and $M_{\rm x} = 6.0 \div 7.7 M_\odot$ for plausible optical component masses in the range $0.09 \div 0.5 M_\odot$.

Orosz (2002) obtained $M_{\rm x} = 6.48 \div 7.19 M_\odot$. 

\tem{(9)} {\bf GS 1124$-$684}

The mass function, given by Orosz et al. (1996), is $f(M_x) = 3.01 \pm 0.15 M_\odot$. In an earlier paper, Orosz et al. (1994) found (from the rotational broadening) $q = 0.133 \pm 0.019$.  Using these data and including the BVRI photometry (ellipsoidal light variations) the authors obtained $i = 54^{\rm o} \div 65^{\rm o}$ and $M_x = 5.0 \div 7.5 M_\odot $.

Later estimates were somewhat more conservative but they did not differ substantially: $f(M_x) = 3.34 \pm 0.15 M_\odot$, $q = 0.128 \pm 0.04$ (Casares et al., 1997), and $i = 39^{\rm o} \div 74^{\rm o}$, $M_x = 3.8 \div 10.5 M_\odot $ (Shahbaz et al., 1997).

The most recent analysis (Gelino et al., 2001a) uses the new JK photometry. They get $i = 54^{\rm o} \pm 2^{\rm o}$ and $M_x = 6.95 \pm 0.6 M_\odot $.

Orosz (2002) obtained $M_{\rm x} = 6.47 \div 8.18 M_\odot$. 

\vspace{6mm}

\tem{(10)} {\bf 4U 1543$-$475}

The first mass function, given by Orosz et al. (1998), was $f(M_x) = 0.22 \pm 0.02 M_\odot$. Orosz et al. analysed also ellipsoidal light variations seen in the V and I light curves. Assuming that the secondary has a normal mass for its spectral type ($\sim 2.3 \div 2.6 M_\odot$), they found $i = 30^{\rm o} \pm 2^{\rm o}$ and $M_{\rm x} \approx 5 \pm 1 M_\odot$.

In the most recent paper, Orosz (2002) quotes for the mass function $f(M_x) = 0.25 \pm 0.01 M_\odot$, for the rotational velocity $v_{rot} sin\, i = 46 \pm 2$ km/sec and for the inclination $i = 20.7^{\rm o} \pm 1.5^{\rm o}$. He concludes that  $M_{\rm x} = 8.45 \div 10.39 M_\odot$. 

\tem{(11)} {\bf XTE J1550$-$564}

The mass function, given by Orosz et al. (2002) is $f(M_{\rm x}) = 6.86 \pm 0.71 M_\odot$. The same authors obtained the tentative measurement of the rotational broadening ($v_{rot} sin\, i = 90 \pm 10$ km/sec) and determined (from the ellipsoidal light variations) the orbital inclination: $i = 73^{\rm o} \pm 2^{\rm o}$. From careful modeling of all observational data, they get $M_{\rm x} = 9.68 \div 11.58 M_\odot$.

Orosz (2002) quotes another solution from Orosz et al. (2002): $M_{\rm x} = 8.36 \div 10.76 M_\odot$. 

\tem{(12)} {\bf GRO J1655$-$40}

The most recent mass function of this famous microquasar was given by Shahbaz et al. (1999): $f(M_{\rm x}) = 2.73 \pm 0.09 M_\odot$. Israelian et al. (1999) measured the rotational broadening: $v_{rot} sin\, i = 93 \pm 3$ km/sec), which implied the mass ratio $q = 0.40 \pm 0.03$. From the careful modeling of the ellipsoidal light variations obtained with the BVIJK photometry, Greene et al. (2001) get $i = 70.2^{\rm o} \pm 1.2^{\rm o}$ and $M_{\rm x} = 6.3 \pm 0.3 M_\odot$. Shahbaz (2002) from precise modeling of the spectrum of the optical component got $q = 0.419 \pm 0.028$ and $M_{\rm x} = 5.99 \pm 0.42 M_\odot$.

Orosz (2002), practically, repeats the solution of Greene et al. ($M_{\rm x} = 6.03 \div 6.57 M_\odot$). 

\tem{(13)} {\bf H 1705$-$250}

The mass function, given by Harlaftis et al. (1997) is $f(M_{\rm x}) = 4.65 \pm 0.21 M_\odot$. The same authors got $q \sim 0.014 (+ 0.019, - 0.012)$ (from the rotational broadening of the optical component lines). Filippenko et al. (1997) obtained  $q \sim 0.01$ from the emission lines of the disc, but they did not consider this result to be realistic. Analysis of the ellipsoidal light variations in B and V bands leads to $i = 70^{\rm o} \pm 10^{\rm o}$ (Remillard et al., 1996), while variations in R band give  $i = 48^{\rm o} \div 51^{\rm o}$ (Martin et al., 1995).
 
Remillard et al., (1996) obtained $M_x = 6 \pm 1 M_\odot $, while Filippenko et al. (1997) got $M_x = 6.4 \div 6.9 M_\odot $ (assuming the mass of the optical component in the range $0.3 \div 0.6 M_\odot $). Harlaftis et al. (1997) state carefully that $M_x = 4.9 \div 7.9 M_\odot $, but their analysis seems to support rather the upper part of this range ($5.7 \div 7.9 M_\odot $).

Orosz (2002) obtained $M_{\rm x} = 5.64 \div 8.30 M_\odot$. 

\tem{(14)} {\bf XTE J1819$-$254}

The mass function, given by Orosz et al. (2001), is $f(M_x) = 2.74 \pm 0.12  M_\odot$. The same authors measured the rotational broadening and found $v_{rot} sin\, i = 124 \pm 4$ km/sec. They analyzed also the ellipsoidal light variations and got $i = 60^{\rm o} \div 71^{\rm o}$. Finally, they concluded that the mass of the compact component is in the range $8.73 \div 11.70 M_\odot$. 

In the most recent paper, Orosz (2002) quotes for the mass function $f(M_x) = 3.13 \pm 0.13 M_\odot$, for the rotational velocity $v_{rot} sin\, i = 98.9 \pm 1.5$ km/sec and for the inclination $i = 75^{\rm o} \pm 2^{\rm o}$. He concludes that $M_{\rm x} = 6.82 \div 7.42 M_\odot$.

\vspace{6mm}

\tem{(15)} {\bf XTE J1859+226}

The preliminary mass function was determined by Filippenko and Chornock (2001): $f(M_x) = 7.4 \pm 1.1 M_\odot$. Ellipsoidal light variations in R band were analyzed by Zurita et al. (2002), who found only that the inclination must be high. The rotational broadening was not measured yet. Assuming that the mass of the optical component is $\ga 1 M_\odot$, one finds $M_{\rm x} \ga 8.0 \div 10.2 M_\odot$.

Orosz (2002) obtained $M_{\rm x} = 7.6 \div 12.0 M_\odot$. 

\tem{(16)} {\bf GRS 1915+105}

The mass function of the celebrated microquasar was determined by Greiner et al. (2001): $f(M_x) = 9.5 \pm 3.0 M_\odot$. The authors assumed then that the inclination is given by the orientation of the jets: $i = 70^{\rm o} \pm 2^{\rm o}$. Assuming, in turn, that the mass of the optical component is $\ga 1.2 M_\odot$, they obtained $M_{\rm x} = 14 \pm 4 M_\odot$. The same value is quoted by Orosz (2002).

\tem{(17)} {\bf GS 2000+251}

The mass function, based on observations with the help of Keck I telescope, is $f(M_x) = 5.01 \pm 0.12 M_\odot $ (Harlaftis et al., 1996). The same observations yielded the estimate of the rotational broadening ($v_{rot} sin\, i = 86 \pm 8$ km/sec), which permitted to determine the mass ratio $q = 0.042 \pm 0.012$. Analysis by Callanan et al. (1996), who included photometry (ellipsoidal light variations) in JK bands and assumed normal (for its spectral type) mass of the secondary ($\sim 0.4 \div 0.9 M_\odot$), gave $i = 65^{\rm o} \pm 9^{\rm o}$ and $M_x = 8.5 
\pm 1.5 M_\odot $. The more recent analysis, incorporating the new J-band photometry (Leeber et al., 1999) gives $i \sim 75^{\rm o}$ and $M_x \sim 6.55 M_\odot $.

Orosz (2002) quotes $i = 64.0^{\rm o} \pm 1.5^{\rm o}$ for the inclination and gets  $M_{\rm x} = 7.15 \div 7.78 M_\odot$. 

\tem{(18)} {\bf GS 2023+338}

The mass function, given by Casares and Charles (1994), is $f(M_x) = 6.08 \pm 0.06 M_\odot $. The same authors analyse the rotational broadening ($v_{rot} sin\, i = 39.1 \pm 1.2$ km/sec), obtaining for the mass ratio the value $q = 0.060 \pm 0.005$. Shahbaz et al. (1994b) included photometry (ellipsoidal light variations) in K band and found $i = 56^{\rm o} \pm 4^{\rm o}$ and $M_x \approx 12 M_\odot$ ($10 \div 15 M_\odot $). Sanwal et al. (1996) analyzed photometry in H band and found that $M_x$ cannot be larger than $\sim 12.5 M_\odot $, and most likely $M_x  \la 9.5 M_\odot $.

Orosz (2002) quotes $i = 38.8^{\rm o} \pm 1.1^{\rm o}$ for the rotational broadening and gets  $M_{\rm x} = 10.06 \div 13.38 M_\odot$.

\section{General Remarks}

\hspace{6mm}$\bullet$ {\bf Brief Statistics}

The list of BHCs (see Table 1) contains 51 objects (including 18 confirmed BHs). About 44 of them (including 13 confirmed) have low mass companions (and so are members of low mass X-ray binaries), while only about 7 (5 confirmed) are members of high mass X-ray binaries. Let us recall, that for the X-ray pulsars (strongly magnetized neutron stars) the distribution is just reversed: out of more than 80 accreting X-ray pulsars, presently known, only few are found in low mass X-ray binaries.

$\bullet$ {\bf Black holes and X-ray Novae}

Galactic binary black holes show clear preference for a specific class of X-ray binaries, namely, for X-ray Novae. This is true also vice versa: great majority of X-ray Novae contain black holes. Out of  $\sim$ 49 XNe known so far (the precise number depends on the definition of an X-ray Nova, so perhaps it is better to talk about SXTs), only in $\sim$ 7 the compact object is a neutron star. The remaining systems ($\sim$ 42) contain BHCs. Since typical XNe remain dormant most of the time, the true number of binary black holes might be easily by one or two orders of magnitude larger than the, 51 objects, listed in Table 1. Most of these black holes reside in presently dormant XNe. Let us recall, that XNe containing BHCs are now being discovered at a rate of 2$-$3 new objects per year.

$\bullet$ {\bf Black holes and microquasars}

Out of 16 known galactic microquasars, two contain neutron stars ( Sco X$-$1 and Cir X$-$1), in one the nature of the compact object is unclear (Cyg X$-$3) and in 13 the compact object is a BHC. Among these 13, there are three superluminal jet sources. There is some evidence that two of superluminal sources contain Kerr (rapidly rotating) BHs.

$\bullet$ {\bf Black holes and stellar evolution}

Analysis of parameters of BHC binaries and their possible evolutionary histories (Kalogera, 1999a,b; Wellstein and Langer, 1999; Brown et al., 1999; Nelemans et al., 1999, Belczy\'nski and Bulik, 2002) leads to interesting conclusions concerning stellar evolution. Among the most important are: (1) BHs are formed more easily than we used to believe; the limiting main sequence mass M$_{\rm NS/BH}$ above which the star ends as a black hole is $\sim$ 20$\div$25 M$_\odot$ rather than $\sim$ 40$-$50 M$_\odot$, as claimed previously; (2) during the supernova explosion of a helium star less than one half of the stellar mass is ejected, (3) the mass loss due to stellar winds from helium stars might be smaller than predicted by canonical formulae.

\section {Acknowledgements}

This work was partially supported by the State Committee for Scientific Research grant No 2 P03C 006 19p01.

%\vspace{15mm}

\section{References}

\ref Abramowicz, M.A. and Klu\'zniak, W.: 2001, \AAL 374 19.
 
\ref Abramowicz, M.A., Bulik, T., Bursa, M. and Klu\'zniak, W.: 2002, \aph 0206490.

\ref Agol, E. and Kamionkowski, M.: 2002, \aph 0109539.

\ref Agol, E., Kamionkowski, M., Koopmans, L. and Blandford, R.: 2002, \aph 0203257.

\ref Antokhina, E.A., Cherepashchuk, A.M.: 1985, {\it Sov. Astr. Lett} {\bf 11}, 4.

\ref Augusteijn, T., Kuulkers, E. and van Kerkwijk, M.H.: 2001, \aph 0107337.

\ref Belczy\'nski, K. and  Bulik, T.: 2002, \aph 0205248.

\ref Belloni, T, Mendez, M., King, A.R., van der Klis, M. and van Paradijs, J.: 1997, \APJL 488 109.

\ref Bennett, D.P., Becker, A.C., Quinn, J.L., Tomaney, A.B., Alcock, C., Allsman, R.A., Alves, D.R., Axelrod, T.S., Calitz, J.J., Cook, K.H., Drake, A.J., Fragile, P.C., Freeman, K.C., Geha, M., Griest, K., Johnson, B.R., Keller, S.C., Laws, C., Lehner, M.J., Marshall, S.L., Minniti, D., Nelson, C.A., Peterson, B.A., Popowski, P., Pratt, M.R., Quinn, P.J., Rhie, S.H., Stubbs, C.W., Sutherland, W., Vandehei, T. and Welch, D.: 2001, \aph 0109467.

\ref Bennett, D.P., Becker, A.C., Calitz, J.J., Johnson, B.R., Laws, C., Quinn, J.L.,  Rhie, S.H. and Sutherland, W.: 2002, \aph 0207006.

\ref Bolton, C.T.: 1972, \NAT 235 271.

\ref Brocksopp, C., Tarasov, A.E., Lyutyi, V.M. and Roche, P.: 1999, \AAP 343 861.

\ref Brown, G.E., Lee, C.-H. and Bethe, H.A.: 1999, \aph 9909270.

\ref
 Callanan, P.J., Garcia, M.R., Filippenko, A.V., McLean, I. And Teplitz, H.: 1996, \APJL 470 57.

\ref
 Callanan, P., McCarthy, J., Garcia, M. and McClintock, J.: 1998, \IAUC 6933.

\ref Casares, J. and Charles, P.A.: 1994, \MN 271 L5.

\ref Casares, J., Martin, E.L., Charles, P.A., Molaro, P. and Rebolo, R.: 1997, {\it New Astronomy} {\bf 1}, 299.

\ref Chen, W., Shrader, C.R. and Livio, M.: 1997, \APJ 491 312.

\ref Cherepashchuk, A.M.: 1998, in {\em Modern Problems of Stellar Evolution}, D.S. Wiebe (ed.),  Geos, Moscow, Russia, p. 198.

\ref Chaty, S. and Mignani, R.P.: 2002, \IAUC 7897.

\ref Chaty, S., Mirabel, I.F., Goldoni, P., Mereghetti, S., Duc, P.-A., Marti, J. and Mignani, R.P.: 2001, \aph 0112329.

\ref Chisholm, J.R., Dodelson, S. and Kolb, E,W.: 2002, \aph 0205138.

\ref Cowley, A.P., Crampton, D., Hutchings, J.B., Remillard, R. and Penfold, J.E.: 1983, \APJ 272 118.

\ref Crampton, D. and Hutchings, J.B.: 1981, \APJ 251 604.

\ref D'Odorico, S., Oosterloo, T., Zwitter, T., Calvani, M.: 1991, \NAT 353 329.

\ref Dolan, J.F. and Tapia, S.: 1989, \APJ 344 830.

\ref Done, C. and Gierli\'nski, M.: 2002, \aph 0211206.

\ref Fabrika, S.N. and Bychkova, L.V.: 1990, \AAL 240 5.

\ref Filippenko, A.V. and Chornock, R.: 2001, \IAUC 7644.

\ref Filippenko, A.V., Matheson, T. and Ho, L.C.: 1995, \APJ 455, 614.

\ref Filippenko, A.V., Matheson, T., Leonard, D.C., Barth, A.J. and Van Dyk, S.D.: 1997, 
\PASP 109 461.

\ref Filippenko, A.V., Leonard, D.C., Matheson, T., Li, W., Moran, E.C. and Riess, A.G.: 1999, 
\PASP 111 969.

\ref Froning, C.S. and Robinson, E.L.: 2001, \AJ 121 2212.

\ref Garcia, M.R., McClintock, J.E., Narayan, R., Callanan, P. and Murray, S.S.: 2001, \APJL 553 47.

\ref Gies, D.R. and Bolton, C.T.: 1982, \APJ 260 240.

\ref Gies, D.R. and Bolton, C.T.: 1986a, \APJ 304 371.

\ref Gies, D.R. and Bolton, C.T.: 1986b, \APJ 304 389.

\ref Gies, D.R., Huang, W. and McSwain, M.V.: 2002, \aph 0208044.

\ref Gelino, D.M., Harrison, T.E. and McNamara, B.J.: 2001a, \AJ 122 971.

\ref Gelino, D.M., Harrison, T.E. and Orosz J.A.: 2001b, \AJ 122 2668.

\ref Greene, J., Bailyn, C.D. and Orosz, J.A.: 2001, \APJ 554 1290.

\ref Greiner, J., Cuby, J.G. and McCaughrean, M.J.: 2001, \NAT 414 522.

\ref Goranskij, V.P.: 1990, {\it IBVS} 3464.

\ref Hanson, M.M., Still, M.D. and Fender, R.P.: 2000, \aph 0005032.

\ref Harlaftis, E.T. and Filippenko, A.V.: 2000, \aph 0002332.

\ref Harlaftis, E.T., Horne, K. and Filippenko, A.V.: 1996, \PASP 108 762.

\ref Harlaftis, E.T., Steeghs, D., Horne, K. and Filippenko, A.V.: 1997, \AJ 114 1170.

\ref Harlaftis, E.T., Collier, S.J., Horne, K. and Filippenko, A.V.: 1999, \AAP 341 491.

\ref Haswell, C.A. and Shafter, A.W.: 1990, \APJL 359 47.

\ref Haswell, C.A., Robinson, E.L., Horne, K., Stiening, R.F. and Abbott, T.M.C.: 1993, \APJ 411 802.

\ref Herrero, A., Kudritzki, R.P., Gabler, R., Vilchez, J.M. and Gabler, A.: 1995, \AAP 297, 556.

\ref Hutchings, J.B., Crampton, D., Cowley, A.P., Bianchi, L. and Thomson, I.B.: 1987, \AJ 94 340.

\ref 
Hynes, R.I., Roche, P. and Haswell, C.A.: 1998, \IAUC 6905.

\ref Ipser, J.R. and Price, R.H.: 1982, \APJ 255 564.

\ref Israelian, G., Rebolo, R., Basri, G., Casares, J. and Martin, E.L.: 1999, \NAT 401 142.

\ref Kalogera, V.: 1999a, \aph 9903417.

\ref Kalogera, V.: 1999b, \aph 9912118.

\ref Kuiper, L., van Paradijs, J. and van der Klis, M.: 1988, \AAP 203 79.

\ref Leeber, D.M., Harrison, T.E. and McNamara, B.J.: 1999, \aph 9911273.

\ref Leibowitz, E.M.: 1984, \MN 210 279.

\ref Lowes, P., in't Zand, J.J.M., Heise, J., Cocchi, M., Natalucci, L., Gennaro, G. and Stornelli, M.: 2002, \IAUC 7843.

\ref Makishima, K., Kubota, A., Mizuno, T., Ohnishi, T., Tashiro, M., Aruga, Y., Asai, K., Dotani, T., Mitsuda, K., Ueda, Y., Uno, S., Yamaoka, K., Ebisawa, K., Kohmura, Y. and Okada, K.: 2000, \APJ 535 632.

\ref Mao, S., Smith, M.C., Wo\'zniak, P., Udalski, A., Szyma\'nski, M., Kubiak, M., Pietrzy\'nski, G., Soszy\'nski, I. and \'Zebru\'n, K.: 2001, \aph 0108312.

\ref Margon, B.: 1984, \ANNREV 22 507.

\ref Markwardt, C.B., Smith, E. and Swank, J.: 2002, \IAUC 7843.

\ref Marsh, T.R., Robinson, E.L. and Woods, J.H.: 1994, \MN 266 137.

\ref Marshall, F.E., Swank, J.H., Thomas, B., Angelini, L., Valinia, A. And Ebisawa, K.: 1997, \IAUC 6543.

\ref  Martin, E.L., Casares, J., Charles, P.A., van der Hooft, F. and van Paradijs, J.: 1995, \MN 274 L46.

\ref Massey, P., Johnson, K.E. and DeDioia-Eastwood, K.: 1995, \APJ 454 151.

\ref McClintock, J.E., Garcia, M.R., Caldwell, N., Falco, E.E., Garnavich, P.M. and Zhao, P.: 2001, \APJL 551 147.

\ref McDovell, J.: 1985, MN 217 77.

\ref McSwain, M.V. and Gies, D.R.: 2002, \APJL 568 27.

\ref 
Mirabel, I.F.: 2000, \aph 0005203.

\ref Narayan, R. and Heyl, J.S.: 2002, \aph 0203089.

\ref Narayan, R., Garcia, M.R. and McClintock, J.E.: 2001, \aph 0107387.

\ref Nelemans, G., Tauris, T.M. and van den Heuvel, E.P.J.: 1999, \aph 9911054.

\ref Orlandini, M., Parmar, A.N., Frontera, F., Masetti, N., Dal Fiume, D., Orr, A., Piccioni, A., Raimondo, G., Santangelo, A., Palazzi, E., Valentini, G. and Belloni, T.: 2000, \AA 356 163.

\ref Orosz, J.A.: 2002, \aph 0209041. 

\ref Orosz, J.A. and Bailyn, C.D.: 1987, \APJ 477 876.

\ref Orosz, J.A., Bailyn, C.D., Remillard, R.A., McClintock, J.E. and Foltz, C.B.: 1994, \APJ 436 848.

\ref Orosz, J.A., Bailyn, C.D., McClintock, J.E. and Remillard, R.A.: 1996, \APJ 468 380.

\ref Orosz, J.A., Jain, R.K., Bailyn, C.D., McClintock, J.E. and Remillard, R.A.: 1998, \APJ 499 375.

\ref Orosz, J.A., Kuulkers, E., van der Klis, M., McClintock, J.E., Jain, R.K., Bailyn, C.D. and Remillard, R.A.: 2000, \IAUC 7440.

\ref Orosz, J.A., Kuulkers, E., van der Klis, M., McClintock, J.E., Garcia, M.R., Callanan, P.J., Bailyn, C.D., Jain, R.K. and Remillard, R.A.:  2001, \APJ 555 489.

\ref Orosz, J.A., Groot, P.J., van der Klis, M., McClintock, J.E., Garcia, M.R., Zhao, P., Jain, R.K., Bailyn, C.D. and Remillard, R.A.: 2002, \APJ 568 845.

\ref Paczy\'nski, B.: 1974, \AAP 34 161.

\ref Paredes, J.M., Mart, J., Ribo, M. and Massi, M.: 2000, {\it Science} {\bf 288}, 2340.

\ref Psaltis, D. and Norman, C.: 2000, \aph 0001391.

\ref Remillard, R.A.: 1999,  \SAIT 70 881.

\ref Remillard, R.A. and Smith, D.A.: 2002, \IAUC 7880.

\ref Remillard, R.A., Orosz, J.A., McClintock, J.E. and Bailyn, C.D.: 1996, \APJ 459 226.

\ref Remillard, R., Levine, A., Swank, J. and Strohmayer, T.: 1997, \IAUC 6710.

\ref Remillard, R., Levine, A. and Wood, A.: 1998, \IAUC 6920.

\ref Remillard, R.A., Muno, M., McClintock, J.E. and Orosz, J.A.: 2002, \aph 0208402.

\ref Remillard, R., Levine, A., Morgan, E.H., Smith, E. and Swank, J.: 2003, \IAUC 8050.

\ref Robinson, E.L., Ivans, I.I. and Welsh,  W.F.: 2002, \APJ 565 1169.

\ref Sanchez-Fernandes, C., Zurita, C. and Casares, J.: 2002, \IAUC 7989.

\ref Sanval, D.,  Robinson, E.L., Zhang, E., Colome, C., Harvey, P.M., Ramseyer, T.F., Hellier, C. and Wood, J.H.: 1996, \APJ 460, 437.

\ref Schmutz, W., Geballe, T.R., Schild, H.: 1996, \AAL 311 25.

\ref Shahbaz, T.: 2002, \aph 0211266.

\ref Shahbaz, T., Naylor, T. and Charles, P.A.: 1994a, \MN 268 756.

\ref Shahbaz, T., Ringwald, F.A., Bunn, J.C., Naylor, T., Charles, P.A. and Casares, J.: 1994b, \MN 271 L10.

\ref Shahbaz, T., Naylor, T. and Charles, P.A.: 1997, \MN 285 607.

\ref Shahbaz, T., van der Hooft, F., Casares, J., T., Charles, P.A. and van Paradijs, J.: 1999, \MN 306 89.

\ref Shrader, C.R. and Titarchuk, L.: 1999, \aph 9906275.

\ref Smith, D.M., Main, D., Marshall, F., Swank, J., Heindl, W.A., Leventhal, M., in't Zand, J.J.M. and Heise, J.: 1998, \APJL 501 181.

\ref Smith, M.C., Mao, S., Wo\'zniak, P., Udalski, A., Szyma\'nski, M., Kubiak, M., Pietrzy\'nski, G., Soszy\'nski, I. and \'Zebru\'n, K.: 2002a, \aph 0206503.

\ref Smith, D.M., Heindl, W.A. and Swank, J.H.:  2002b, \aph 0209255.

\ref Sunyaev, R. and Revnitsev, M.: 2000, \aph 0003308.

\ref Tanaka, Y., Lewin, W.H.G.: 1995, in {\it X-ray Binaries}, W.H.G. Lewin, J. van Paradijs and E.P.J. van den Heuvel (eds), Cambridge Univ. Press, Cambridge, U.K., p. 126.

\ref Taylor, A.R., Gregory, P.C., Duric, N.,Tsutsumi, T.: 1991, \NAT 351 547.

\ref Thorsett, S.E. and Chakrabarty, D.: 1999, \APJ 512 288.

\ref Tomsick, J.A., Kalemci, E., Corbel, S. and Kaaret, P.: 2002, \IAUC 7837.

\ref Vacca, W.D., Garmany, C.D. and Shull, J.M.: 1996, \APJ 454 151.

\ref van Paradijs, J.: 1998, \aph 9802177.

\ref Wagner, R.M., Foltz, C.B., Shahbaz, T., Casares, J., Charles, P.A., Starrfield, S.G. and Hewett, P.: 2001, \APJ 556 42.

\ref Wagoner, R.V., Silbergleit, A.S. and Ortega-Rodriguez, M.: 2001, \APJL 559 25.

\ref Webb, N.A., Naylor, T., Ioannou, Z., Charles, P.A. and Shahbaz, T.: 2000, \MN 317 528.

\ref Wellstein, S. and Langer, N.: 1999, \aph 9904256.

\ref Zdziarski, A.A., Poutanen, J., Miko{\l}ajewska, J., Gierli\'nski, M., Ebisawa, K. and Johnson, W.N.: 1998, \MN 301 435.

\ref Zhang, W., Smale, A.P., Strohmayer, T.E., Swank, J.H. 1998, \APJL 500 171.

\ref Zi\'o{\l}kowski, J.: 2002, in {\it Proceedings of the 4-th Microquasar Workshop}, Ph. Durouchoux, Y. Fuchs and J. Rodriguez (eds), Center for Space Physics: Kolkata (in press).

\ref Zurita, C., Sanchez-Fernandes, C., Casares, J., Charles, P.A., Abbott, T.M., Hakala, P., Rodriguez-Gil, P., Bernabei, S., Piccioni, A., Guarnieri, A., Bartolini, C.,  Masetti, N., Shahbaz, T. and Castro-Tirado, A.: 2002, \aph 0204337.

\end{document}